\documentclass[a4paper]{aa}
\usepackage[T1]{fontenc}
\usepackage[utf8]{inputenc}
\usepackage{graphicx}
\usepackage{color}
\usepackage{txfonts}
\usepackage{microtype}
\usepackage{ellipsis}
\usepackage{natbib}

\bibpunct{(}{)}{;}{a}{}{,}
\usepackage[pdftex,colorlinks=true,urlcolor=blue]{hyperref}
\usepackage[xindy,nonumberlist,nopostdot,nogroupskip]{glossaries}

\makeglossaries
\usepackage[xindy]{imakeidx}
\makeindex

\begin{document}
\title{The Intertropical Convergence Zone}
\author{M. Nielbock}
\institute{Haus der Astronomie, Campus MPIA, Königstuhl 17, D-69117 
Heidelberg, Germany\\
\email{nielbock@hda-hd.de}}

\date{Received February 18, 2016; accepted }

\abstract{This activity has been developed as a resource for the ``EU Space Awareness'' educational programme. As part of the suite ``Our Fragile Planet'' together with the ``Climate Box'' it addresses aspects of weather phenomena, the Earth's climate and climate change as well as Earth observation efforts like in the European ``Copernicus'' programme. This resource consists of three parts that illustrate the power of the Sun driving a global air circulation system that is also responsible for tropical and subtropical climate zones. Through experiments, students learn how heated air rises above cool air and how a continuous heat source produces air convection streams that can even drive a propeller. Students then apply what they have learnt to complete a worksheet that presents the big picture of the global air circulation system of the equator region by transferring the knowledge from the previous activities in to a larger scale.}


\maketitle
%
%________________________________________________________________

\section{Background information}
\subsection{Temperature, density and buoyancy}
We all know from experience that the\newglossaryentry{density}
{name = Density,
description = {The density is a measure of the specific weight or mass of a given substance or object. A higher density results in a higher weight or mass for the same amount or volume of that substance. Mathematically, this is the mass divided by the volume.}}
density of gases depends on their temperature. Cold air is denser than warm air. This is the reason why hot air balloons can fly and even lift up cargo.

\begin{figure}
\centering
\resizebox{\hsize}{!}{\includegraphics{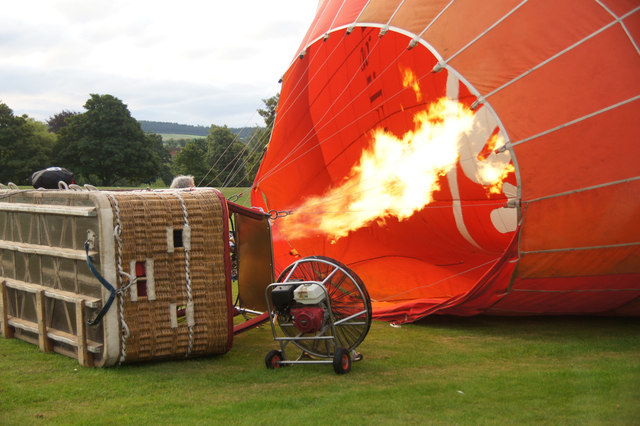}}
\caption{Heating and inflating a hot-air balloon (Mike Pennington, \url{http://www.geograph.org.uk/photo/3605709}, \url{https://creativecommons.org/licenses/by-sa/2.0/legalcode}).}
\label{f:balloon}
\end{figure}

Scientifically, this phenomenon is described by the Ideal Gas Law\newglossaryentry{IDG}
{name = {Ideal Gas Law},
description = {The ideal gas law is the equation of state of a hypothetical ideal gas, usually expressed by thermodynamical quantities like pressure, temperature, volume, mass or particle number. It is a good approximation of the behaviour of many gases under many conditions.}}
and the buoyant\newglossaryentry{buoyancy}
{name = Buoyancy,
description = {An upward force exerted by a gas or a fluid that opposes the weight of an immersed object.}}
force according the Archimedes' principle.\newglossaryentry{archimedes}
{name = {Archimedes' principle},
description = {The Archimedes' principle states that the upward buoyant force that is exerted on a body immersed in a fluid, whether fully or partially submerged, is equal to the weight of the fluid that the body displaces and acts in the upward direction at the centre of mass of the displaced fluid.}}
Hence, the buoyant force is equivalent to the gravitational force that is exerted on the displaced air while the balloon inflates.
\begin{equation}
F_{\rm up} = F_{\rm g,air,out} = m_{\rm air,out} \cdot g
\end{equation}
Here $m_{\rm air,out}$ is the mass of the displaced air. In case of hot air balloons with the volume $V$ and the density of the air outside the balloon that was displaced $\varrho_{\rm air,out}$:
\begin{equation}
F_{\rm up} = \varrho_{\rm out} \cdot V\cdot g
\end{equation}
Buoyancy is reached, when the buoyant force is equivalent to the gravitational force (only the air is considered):
\begin{equation}
F_{\rm up} = F_{\rm g,air,in} = m_{\rm air,in} \cdot g = \varrho_{\rm in} \cdot V \cdot g
\end{equation}
Hence:
\begin{eqnarray}
F_{\rm res} &=& F_{\rm up} - F_{\rm g,air,in}\\
            &=& \varrho_{\rm out} \cdot V \cdot g - \varrho_{\rm in} \cdot V \cdot g\\
            &=& \left(\varrho_{\rm out} - \varrho_{\rm in}\right) \cdot V \cdot g
\end{eqnarray}
So, the resulting force is governed by the difference between the densities of the air inside and outside the balloon. For a hot air balloon, this density inside is changed by heating the air. This is where the Ideal Gas Law steps in. One way of writing it is:
\begin{equation}
p\cdot V=m\cdot R_s\cdot T
\end{equation}
$m$ is the total gas mass and $R_s$ is a constant. Therefore:
\begin{equation}
\frac{p\cdot V}{m\cdot T}=\frac{p}{\varrho \cdot T} = {\rm const.}
\end{equation}
Since a hot air balloon is open, the pressure inside and outside the balloon remains the same during the heating process. It is a constant as well. Therefore:
\begin{equation}
\varrho \cdot T = {\rm const.}
\end{equation}
When beginning heating the air inside the balloon, both the density and the temperature inside and outside are the same. Afterwards both quantities differ. Therefore, we can write:
\begin{equation}
\varrho_{\rm in}\cdot T_{\rm in} = \varrho_{\rm out}\cdot T_{\rm out} \Leftrightarrow \varrho_{\rm in} = \varrho_{\rm out} \cdot \frac{T_{\rm out}}{T_{\rm in}}
\end{equation}
Let us now get back to the resulting force of uplift:
\begin{equation}
F_{\rm res} = \left(\varrho_{\rm out} - \varrho_{\rm in}\right) \cdot V \cdot g
\end{equation}
The two densities are now connected to the temperature change. Therefore:
\begin{eqnarray}
\varrho_{\rm out} - \varrho_{\rm in} &=& \varrho_{\rm out} - \varrho_{\rm out} \cdot \frac{T_{\rm out}}{T_{\rm in}}\\
                                     &=& \varrho_{\rm out} \cdot \left(1-\frac{T_{\rm out}}{T_{\rm in}}\right)
\end{eqnarray}
Which leads to final expression for the resulting force of uplift:
\begin{equation}
F_{\rm res} = \varrho_{\rm out} \cdot \left(1-\frac{T_{\rm out}}{T_{\rm in}}\right) \cdot V \cdot g
\end{equation}
In order for $F_{\rm res}$  to be positive, we need:
\begin{equation}
1-\frac{T_{\rm out}}{T_{\rm in}} > 0 \Leftrightarrow T_{\rm in} > T_{\rm out}
\end{equation}
So, the hot air balloon (in fact only the air inside it) rises, if the air temperature inside the balloon is higher than outside. In reality, the balloon lifts up, if the resulting force also compensates the gravitational force i.e. the weight of the balloon itself. As a result, a balloon of 18~m in diameter produces an uplift equivalent to a metric ton, if the air inside is heated from $15\degr$C to $120\degr$C, the maximum allowed for nylon fabric.

\subsection{The Sun and the global wind system}
\begin{figure}
\centering
\resizebox{\hsize}{!}{\includegraphics{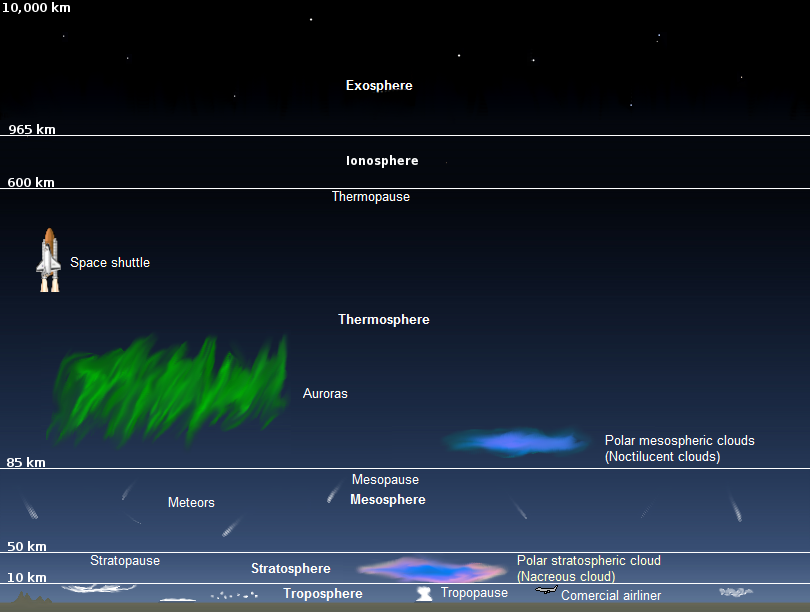}}
\caption{Layers of the atmosphere (The High Fin Sperm Whale, \url{https://commons.wikimedia.org/wiki/
File:Layers_of_the_atmosphere.PNG}, ``Layers of the atmosphere'', altitudes according to NASA (\url{https://www.nasa.gov/
mission_pages/sunearth/science/atmosphere-layers2.html}) added by M. Nielbock, \url{https://creativecommons.org/
licenses/by-sa/3.0/legalcode}).}
\label{f:layers}
\end{figure}

The atmosphere of the Earth consists of air. We know that whenever it is heated, it rises up (see above). Heating of the atmosphere is done by the Sun, mainly indirectly via heating the surface first which by producing infrared radiation in turn heats the air close to the ground. Since warm air has a lower density than cool air, it produces an updraft which by convection drags the air up into the highest layers of the troposphere about 10~--~15~km above ground (Fig.~\ref{f:layers}). While the air rises, it constantly cools down which reduces the buoyant force. At some point, the uplift stops, and the air is diverted horizontally north and south.

This process is most effective at latitudes where the Sun is at zenith, i.e.~in a belt around the equator. This area is called the Intertropical Convergence Zone (ITCZ). It follows the apparent annual path of the Sun northward and southward during the seasons.

Cool air can store less water than warm air. Therefore, the rather humid air that rises in the ITCZ continuously loses its ability to store the water. As a result, clouds form from which eventually water is released as rain. In extreme cases, the convection can lead to severe weather phenomena like thunderstorms or even cyclones and hurricanes. This is the reason, why the low latitudes around the equator are strongly affected by a humid, tropical climate with low pressure areas and lots of rain. On satellite images, the ITCZ is very prominent due to a global cloud belt along the equator (see Figure 1). This demonstrates that satellite imagery can be an important tool to monitor the climate of the Earth.

\begin{figure}
\centering
\resizebox{\hsize}{!}{\includegraphics{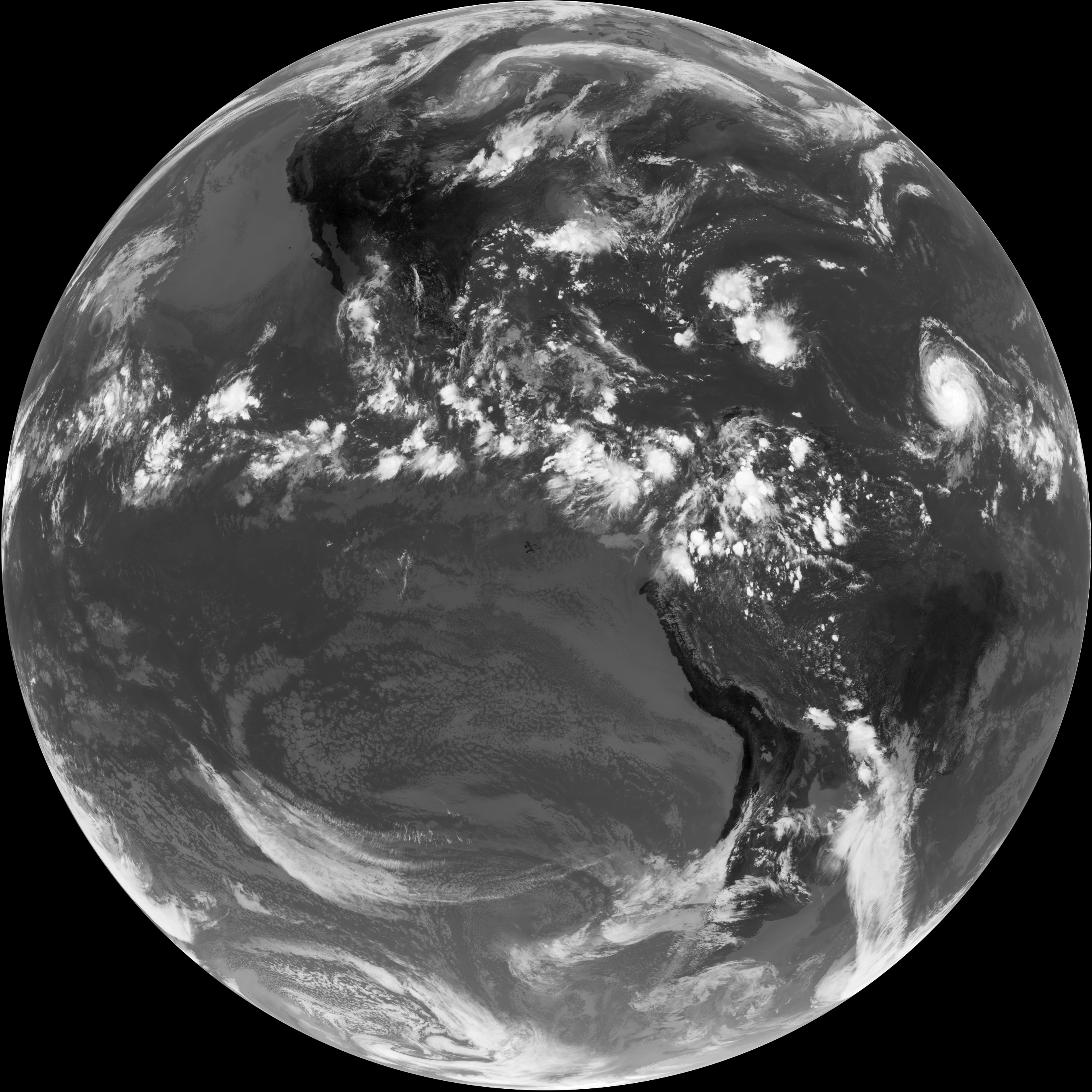}}
\caption{Image obtained with the GOES 14 satellite. The belt of cloud formation around the ITCZ is clearly visible (NASA/GOES Science Team).}
\label{f:satimage}
\end{figure}

The air has now become dry. While it is pushed north and south to latitudes around $30\degr$ north and south, the already dried air drops, heats up, and is dried even more. This gives rise to the arid\newglossaryentry{arid}{name=Arid,description={A region is arid when it is characterised by a severe lack of available water, to the extent of hindering or preventing the growth and development of plant and animal life.}}, subtropical climate we find there.

\begin{figure}
\centering
\resizebox{\hsize}{!}{\includegraphics{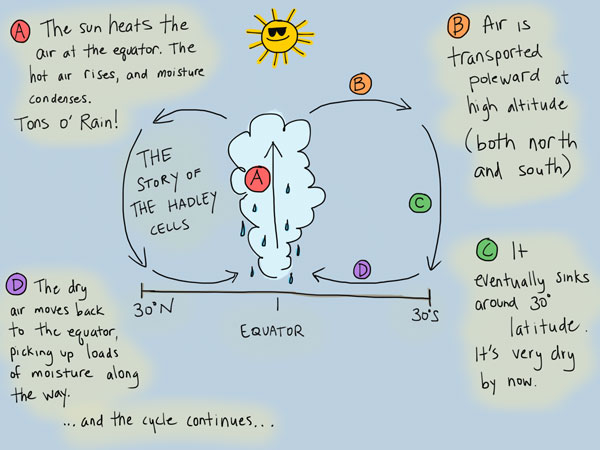}}
\caption{: Schematic of the Hadley cell (\href{http://www.michw.com/}{``The Waveform Diary'' blog}, \href{http://www.michw.com/2013/05/the-mystery-of-the-shifting-tropical-rain-belt/}{``The mystery of the shifting tropical rain belt''}, M. Weirathmueller, permission for reproduction granted).}
\label{f:hadley}
\end{figure}

The cycle of the circulation closes with the air currents flowing back to the equatorial region and the ITCZ where they feed the convection again. Altogether, we have seen that the ITCZ with the rising air causes a global belt of air circulating roughly between the tropics around the equator and the dry deserts at approximately $30\degr$ north and south. This belt is called the \newglossaryentry{hadley}{name={Hadley cell},description={The Hadley cell, named after George Hadley, is a global scale tropical atmospheric circulation that features air rising near the equator, flowing poleward at 10--15 kilometres above the surface, descending in the subtropics, and then returning equatorward near the surface. This circulation creates the trade winds, tropical rain-belts and hurricanes, subtropical deserts and the jet streams.}}``Hadley cell'' (see Fig.~\ref{f:hadley}).

\begin{figure}
\centering
\resizebox{\hsize}{!}{\includegraphics{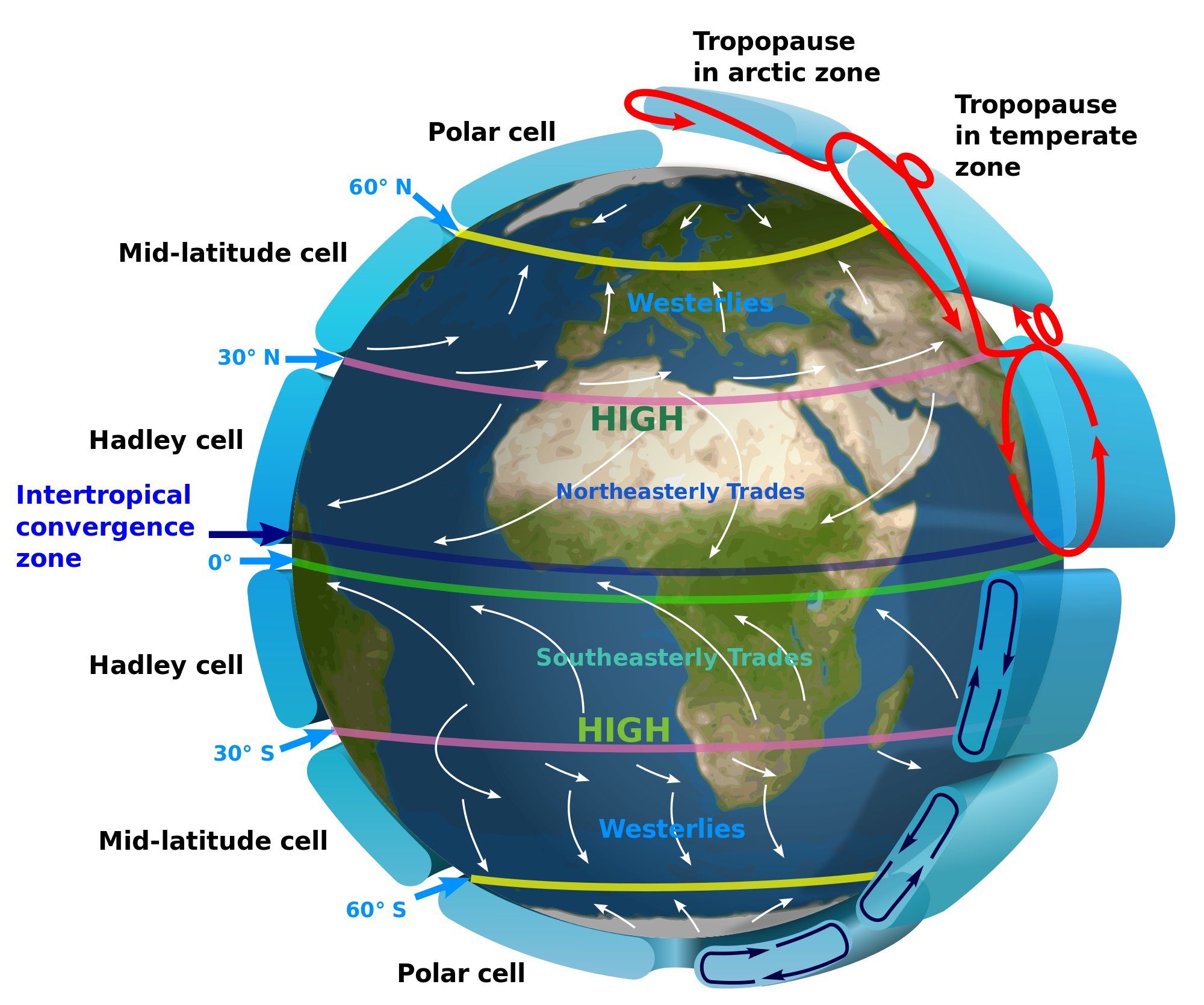}}
\caption{Global circulation of Earth's atmosphere (Kaidor, \url{https://commons.wikimedia.org/wiki/
File:Earth_Global_Circulation_-_en.svg}, ``Earth Global Circulation - en'', \url{https://creativecommons.org/
licenses/by-sa/3.0/legalcode}).}
\label{f:winds}
\end{figure}

The convection zone at the ITCZ and the connected Hadley zone is only a part of a much larger and global air circulation system (see Fig.~\ref{f:winds}).

\subsection{Possible influence of climate change}
We have seen that thermal energy is the driver of wind, rain and temperature, i.e.~the weather. Therefore, it seems plausible that the more heat is stored inside the atmosphere, the more energy is sitting there to influence the weather. Since the resulting buoyant force depends on temperature differences, especially between the Earth's surface and high altitudes, we should expect that processes as those that we see inside the ITCZ are amplified.

In fact, there are already indications that along with the phenomenon of a gradual increase of global mean air temperatures throughout the recent decades (global warming), extreme weather situations have increased in number and severity (\url{http://www.ucsusa.org/global_warming/science_and_impacts/impacts/global-warming-rain-snow-tornadoes.html}). This is supported by recent results in climate results (\url{https://www.sciencedaily.com/releases/2017/03/170307100337.htm}). 

\section{List of material}
The activities are carried out best in groups of two. The items listed below are indicated per group.

\medskip\noindent
Common items:
\begin{itemize}
\item paper handkerchief, napkin or dual chamber tea bag
\item matches or lighter
\item china plate or tinfoil (or any other non-flammable flat surface)
\item strong lamp (common bulb or tungsten halogen bulb, min.~100~W)
\item scissors
\item flat nose pliers, if available
\item glue (for cardboard)
\item pencil or similar pointed object
\item aluminium wrap of a tea light
\item drawing pin 
\end{itemize}

\medskip\noindent
For either of the following alternatives:

\medskip\noindent
{\em Alternative 1}
\begin{itemize}
\item cardboard tube (inner part of a kitchen roll)
\item black paint and brush or black coloured paper
\item one piece of cardboard (approx.~1 cm~wide, 8~cm long)
\end{itemize}

\noindent
{\em Alternative 2}
\begin{itemize}
\item construction template provided with this sheet
\item black cardboard (22 cm x 20 cm)
\item one piece of card board (approx. 1cm wide, 12 cm long, see template)
\end{itemize}

\section{Goals}
Students will learn how the Sun drives the main engine of atmospheric circulation. They will experience hands-on, how irradiation and heat can drive convection.  Students learn that warm air rises over cold air. They will understand that this basic phenomenon is the cause for the large scale air circulation systems on Earth and the warm and humid climate near the equator.

\section{Learning objectives}
{\em Desired Student Outcomes for Activity 1:}
After this activity, the students will be able to explain that warm air rises above the surrounding cooler air and that the uplift can be strong enough to launch objects up into the air.

\medskip\noindent
{\em Desired Student Outcomes for Activity 2:}
After this activity, the students will be able to explain that a continuous convection stream can be produced, provided the energy source injects heat constantly. They will also understand that a supply of fresh air is needed to keep the engine going.

\medskip\noindent
{\em Desired Student Outcomes for Activity 3:}
This worksheet helps students to transfer the knowledge learnt in the first two activities to the situation on Earth, where the Sun is the main heat source. Students answer questions on how the air on Earth is heated and the effects of this. They will understand that the Intertropical Convergence Zone is only a part of a larger air circulation system that is the cause for the climate zones between the Tropics and the Subtropics.

\section{Target group details}

\noindent
Suggested age range: 12 -- 16 years\\
Suggested school level: Middle School, Secondary School\\
Duration: 1 hour

\section{Evaluation}
The answers given by the students to the questions as listed in the activity -- partially stimulated by discussion -- are the gauge by which the teacher can evaluate the learning outcome.

The teacher could ask students to draw a labelled diagram for the first two activities to show what they saw and write a couple of sentences about what happened and why.

The final activity, a worksheet, summarises what has been illustrated and discussed in the first two parts and puts it in the perspective of a global system on Earth. From this, it should be straightforward to judge to what degree the individual elements have been understood.

For a solution for the drawing, see Fig.~\ref{f:hadley} in the Background information.

\section{Full description of the activity}
\subsection{Introduction: Questions and Answers}
Show the students an Earth globe or a global map and ask them if they can identify the equator.

\medskip\noindent
Q: Where are the poles? Where does the equator lie relative to the poles?\\\noindent
A: The equator is the line or circumference on the globe half way between the poles.

\begin{figure}[!ht]
\centering
\resizebox{0.6\hsize}{!}{\includegraphics{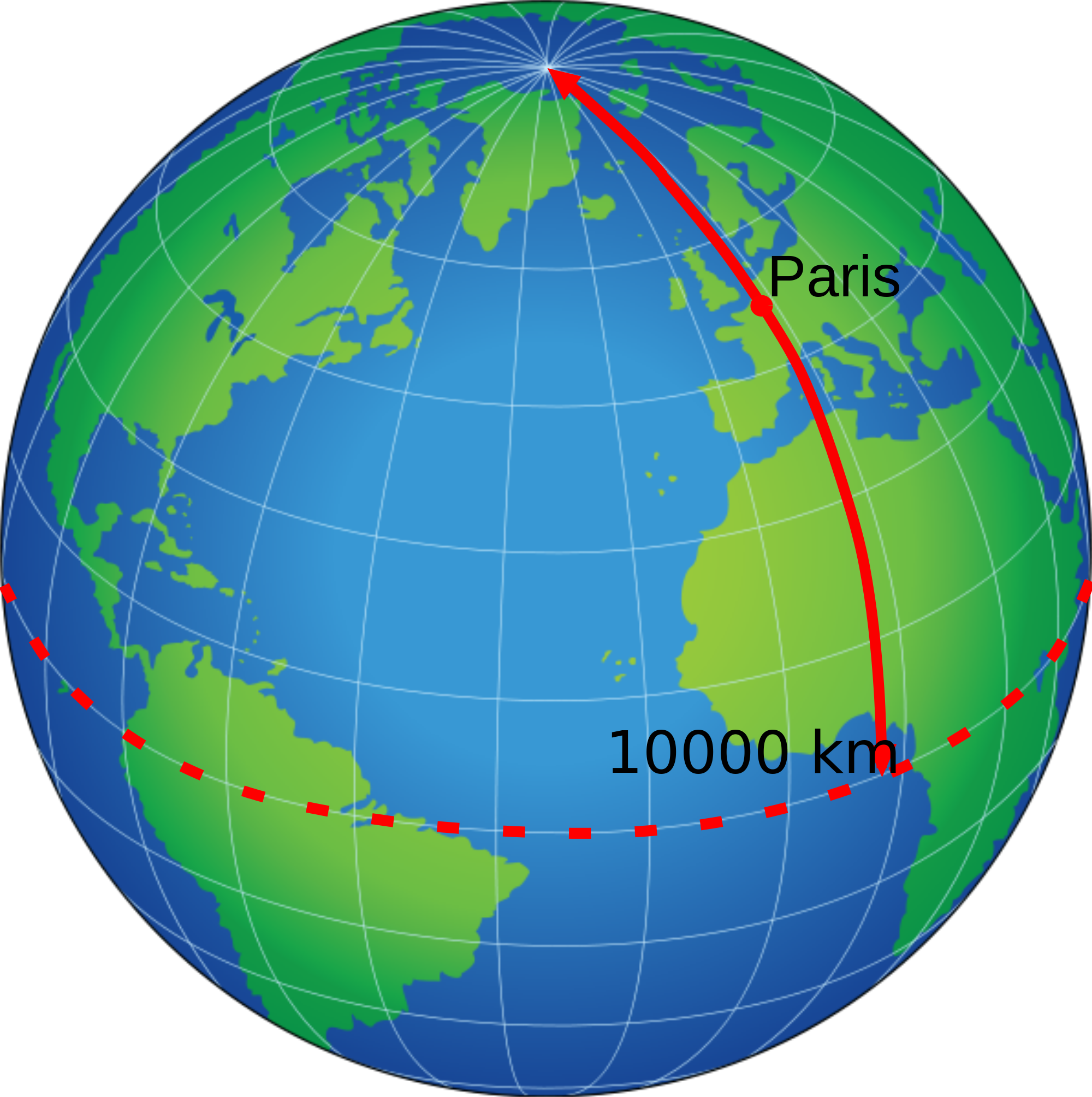}}
\caption{Illustration of the Earth globe and the position of the equator relative to the North Pole (US Government).}
\label{f:globe}
\end{figure}

If possible, let the students use an on-line map tool to investigate the geography and satellite images of the equator region. The map tool may also be projected by one computer only. Possible map services with satellite imagery:

\medskip\noindent
\url{https://maps.google.com}\\\noindent
\url{https://www.arcgis.com/home/webmap/viewer.html?useExisting=1}

\medskip\noindent
Q: Can you name countries that touch the equator? Do you know any cities close to it?\\\noindent
A: e.g. Ecuador (Quito, Galapagos), Brazil, Gabon (Libreville), Congo, Uganda (Kampala), Kenya, Malaysia, Indonesia.

\medskip\noindent
Q: What is the typical vegetation there? A closer look at the satellite images helps to answer the question.\\\noindent
A: Rainforest.

\medskip\noindent
Q: What is the typical weather there (humid or dry, cold or warm)?\\\noindent
A: Humid and warm, lots of rain.

\begin{figure}[!ht]
\centering
\resizebox{\hsize}{!}{\includegraphics{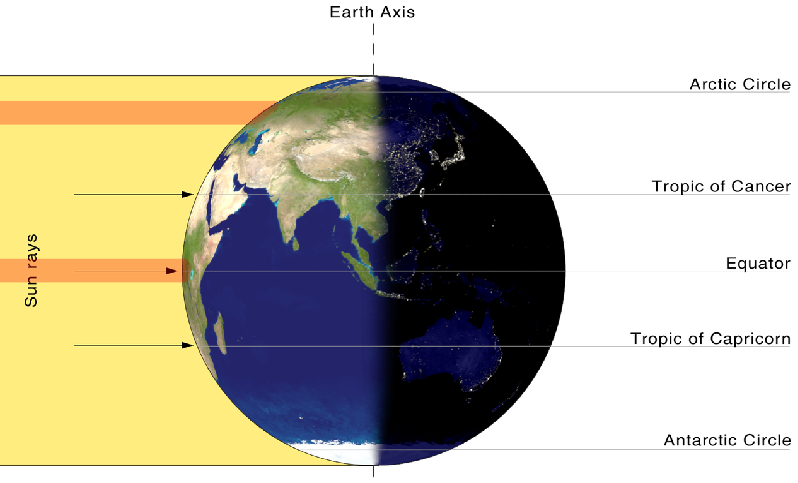}}
\caption{Illustration of the Earth by the Sun. Looking from the equator, the Sun is almost always directly above (Przemyslaw ``Blueshade'' Idzkiewicz, \url{http://commons.wikimedia.org/wiki/File:Earth-lighting-equinox\_EN.png}, ``Earth-lighting-equinox EN'', solar light beams at different latitudes added by Markus Nielbock, \url{https://creativecommons.org/licenses/by-sa/2.0/legalcode}).}
\label{f:irradiation}
\end{figure}

\medskip\noindent
Q: Can you think of a reason, why it is so warm there all year? Show them a model of the Sun-Earth-System (Fig.~\ref{f:irradiation}).\\\noindent
A: The Sun is the main heating source. Near the equator, it is always almost directly above.

\noindent
Q: The air is heated up by the hot surface of the Earth. What happens with hot air? Imagine a hot-air balloon.\\\noindent
A: Warm/hot air rises above cold air.

\medskip\noindent
Q: How would you be able to recognise that the air rises?\\\noindent
A: Wind

\subsection{Activity 1: Flying flames}
\textbf{WARNING! This activity is only suitable for students that are responsible enough to handle a flame.} Be sure to provide an inflammable environment and to supervise the students during the experiment. It is suggested to distribute lighters or matches only to small groups of students at a time.  If in doubt, the experiment should be demonstrated by the teacher.

Smoke detectors may have to disabled for this.

Use a fireproof surface. A china plate or a piece of tinfoil may be used to protect the surface of normal desks or tables.

In this activity, the students will experience, how warm air rises above cooler air. The hot air produced by a burning piece of very light paper produces its own uplift and rises up in the air. This experiment should give rather a qualitative result meaning that the exact uplift force is not so important.

Gather the following items, one set per group (or one for the teacher only, if carried out as a demonstration):
\begin{itemize}
\item paper handkerchief, napkin or dual chamber tea bag
\item matches or lighter
\item plate or/and tinfoil
\item scissors
\end{itemize}

\begin{figure}[!ht]
\centering
\resizebox{\hsize}{!}{\includegraphics{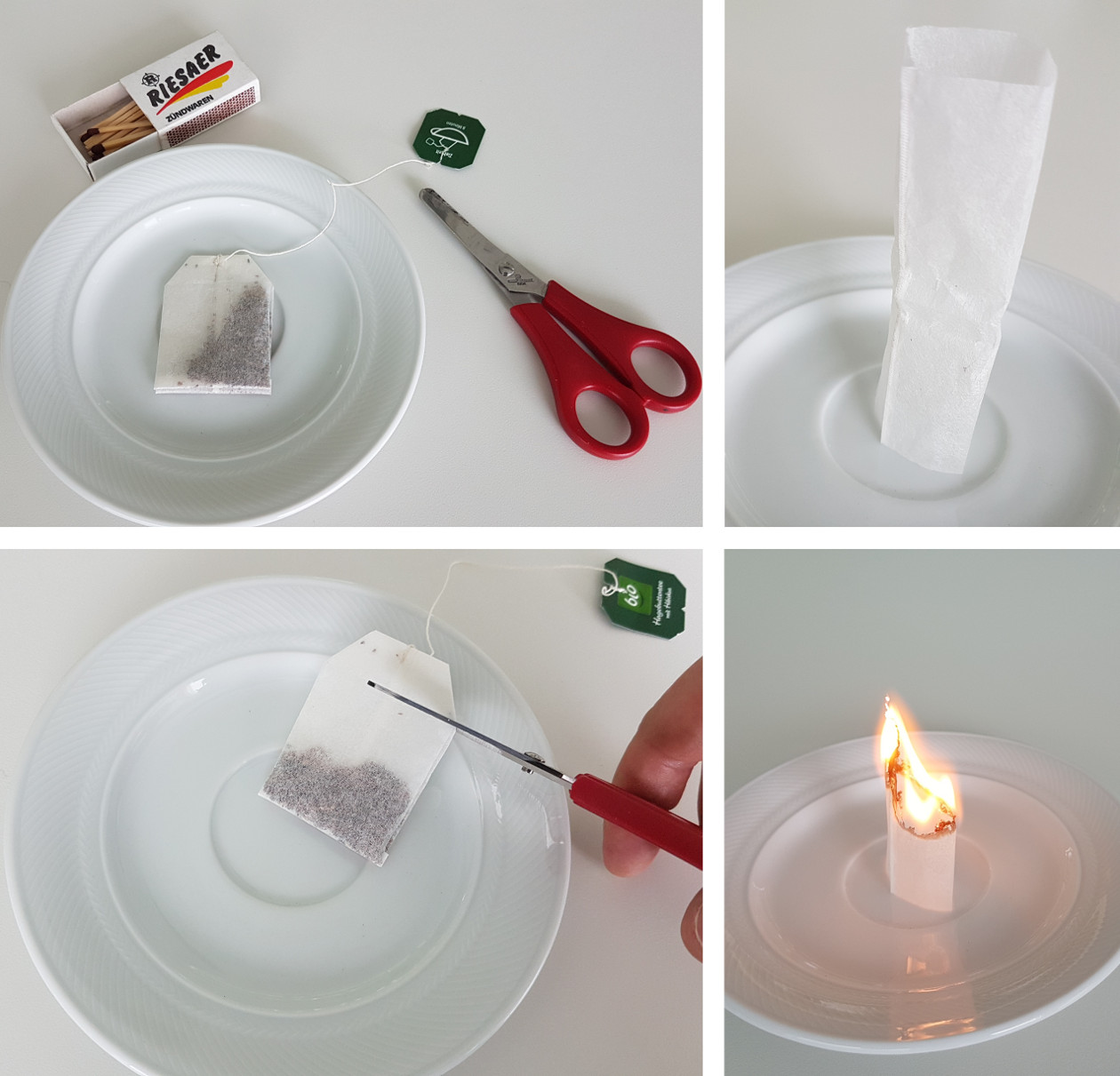}}
\caption{Illustration of the instructions (own work).}
\label{f:a1}
\end{figure}

Distribute the students in groups of two (suggested).
\begin{enumerate}
\item Prepare the wick:
 \begin{enumerate}
 \item Paper handkerchiefs and napkins consist of several layers. Take only one and cut off one quarter.
 \item If a tea bag is used, cut off the top and empty the bag. Unfold it.
 \end{enumerate}
 \item Form a tube (napkin, handkerchief, tea bag) of a few centimetres and put it on the plate, standing upright. It should stand stably. Avoid abrupt and fast movements to prevent moving air from blowing the wick away.
 \item Light it.
\end{enumerate}

Discuss with the students what happened. Let the students describe in detail what they saw. While the wick burns down, it lifts off at some point.

\medskip\noindent
Q: What happened to the air around the burning paper?\\\noindent
A: It was heated.

\medskip\noindent
Q: What happens with heated air?\\\noindent
A: It rises.

\medskip\noindent
Q: Can you explain why in the end the burning paper lifted off?\\\noindent
A: It was dragged along with the rising air.

\subsection{Activity 2: Updraft tower}
This activity demonstrates how heated air rises. As long as the heating source is present, a continuous updraft of the heated air is generated. The students will experience this phenomenon by building a model of an updraft tower. Afterwards, the concept of air circulation can be used to explain the process of terrestrial atmospheric circulation systems and the Intertropical Convergence Zone.
\newglossaryentry{sut}{name={Solar updraft tower},description={The solar updraft tower (SUT) is a renewable-energy power plant for generating electricity from low temperature solar heat. Sunshine heats the air beneath a very wide greenhouse-like roofed collector structure surrounding the central base of a very tall chimney tower.}}

Tell the students that they will now build a model of such a tower. Tell them that it is a nice example of demonstrating vertical wind on a small scale. Gather the following items, one set per group.

\begin{itemize}
\item scissors
\item flat nose pliers, if available
\item glue (for cardboard)
\item pencil or similar pointed object
\item aluminium wrap of a tea light
\item drawing pin 
\end{itemize}

\medskip\noindent
For either of the following alternatives:

\medskip\noindent
{\em Alternative 1}
\begin{itemize}
\item cardboard tube (inner part of a kitchen roll)
\item black paint and brush or black coloured paper
\item one piece of cardboard (approx.~1 cm~wide, 8~cm long)
\end{itemize}

\noindent
{\em Alternative 2}
\begin{itemize}
\item construction template provided with this sheet
\item black cardboard (22 cm x 20 cm)
\item one piece of card board (approx. 1cm wide, 12 cm long, see template)
\end{itemize}

\begin{figure}
\centering
\resizebox{\hsize}{!}{\includegraphics{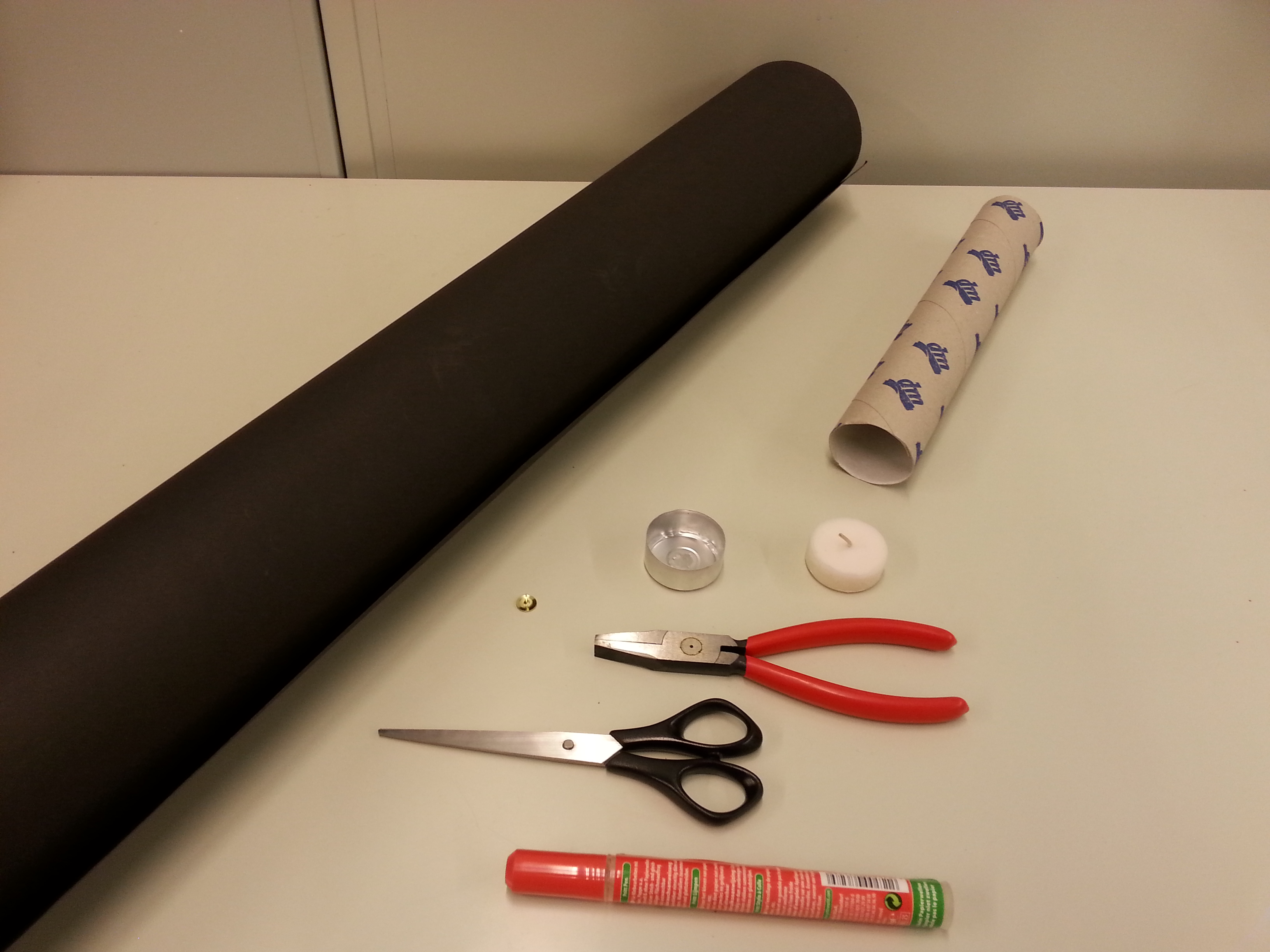}}
\caption{Items needed for building an updraft tower model (own work).}
\label{f:a2_1}
\end{figure}

\subsubsection{Building instructions}
\begin{figure}[!ht]
\centering
\resizebox{0.6\hsize}{!}{\includegraphics{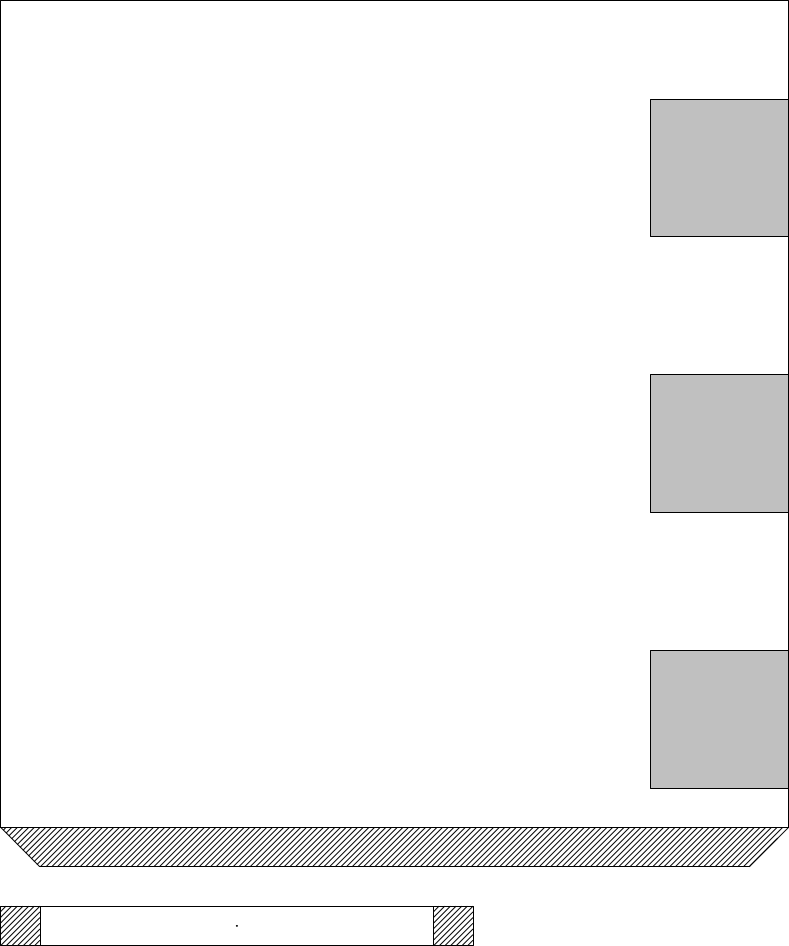}}
\caption{Scaled down version of the construction template. The proper version is available as a separate file (own work).}
\label{f:a2_template}
\end{figure}

\begin{figure*}
\centering
\resizebox{\hsize}{!}{\includegraphics{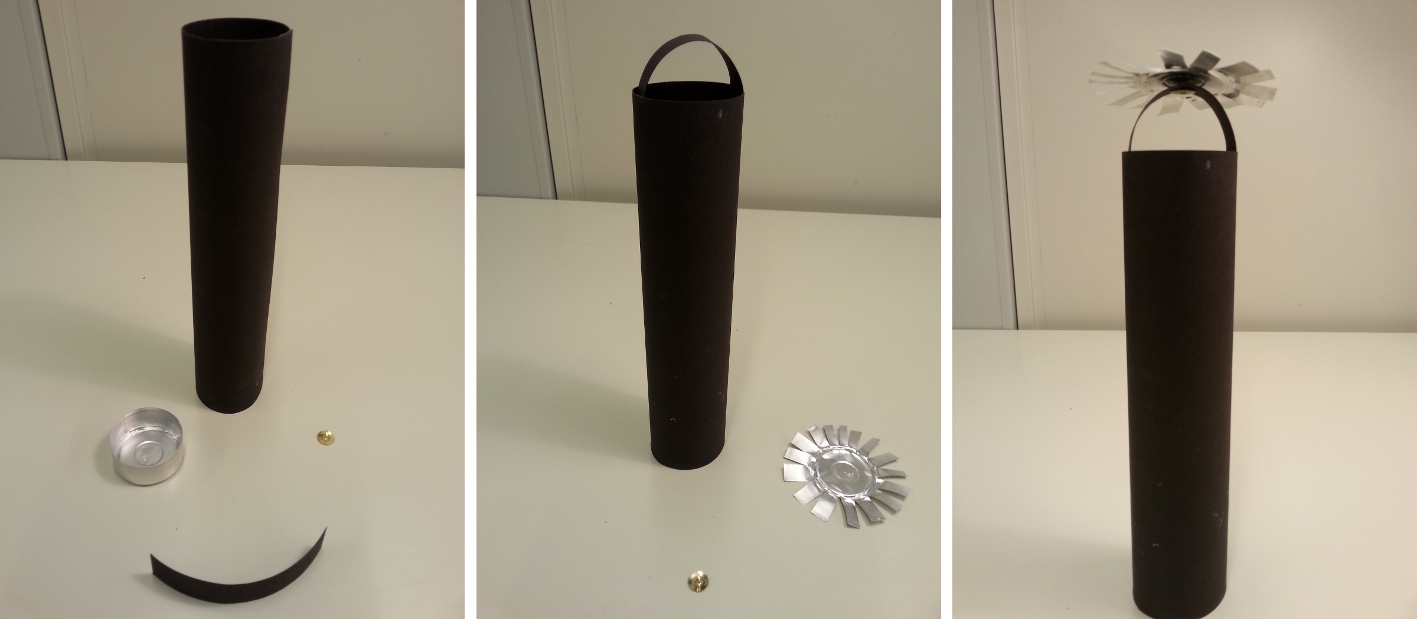}}
\caption{Various steps while building the updraft tower (own work).}
\label{f:a2_2}
\end{figure*}

\noindent
{\em Alternative 1:}\\\noindent
The first version may be somewhat simpler to produce, but might not be that effective, because the cardboard tube used in this example may a bit too narrow.

\begin{enumerate}
\item Paint the outside of the cardboard tube black or glue it with black paper.
\end{enumerate}

\noindent
{\em Alternative 2:}\\\noindent
The second version is especially designed to match the diameters of the tower and the fan.

\begin{enumerate}
\item Prepare the black cardboard according to the construction template provided.
\item Roll the cardboard perpendicularly to the hashed area.
\item Glue the tube at the hashed area.
\item Cut out the grey areas or cut them from bottom to top and fold them up to form flaps (Fig.~\ref{f:a2_2}, left).
\end{enumerate}

\noindent
{\em Common steps:}
\begin{enumerate}
\setcounter{enumi}{4}
\item Now we produce the fan using the tea light wrap. This part is quite delicate and has to be done very carefully.
\item Cut 16 equal sections into the walls of the tea light wrap.
\item Flatten the sections outside to the bottom of the wrap.
\item Extend the cuts to the inner circle of the bottom of the wrap.
\item Press the pencil exactly at the centre of the fan to form a small dent. Be careful not to punch a hole.
\item Bend all 16 wings of the fan around an axis from the centre to the edge. Use the pliers if available.
\item Punch the drawing pin through the centre and from the back of the small piece of cardboard.
\item Glue the small piece of cardboard to the inside at the top of the tube. It should form an arc (Fig.~\ref{f:a2_2}, middle).
\item Put the fan on top of the drawing pin.
\item Balance the fan by bending the wings up and down.
\item Illuminate the side of the tower with a strong lamp.
\item Watch the fan rotate (Fig.~\ref{f:a2_2}, right).
\end{enumerate}

\noindent
Discuss the results with the students.

\medskip\noindent
Q: Why does the fan rotate?\\\noindent
A: The air inside the tower is heated up and streams upward.

\medskip\noindent
Q: How is the air heated? Remember that the lamp does not shine inside the tower.\\\noindent
A: The lamp heats the black tower, which in turn heats the air inside.

\medskip\noindent
Q: What are holes at the bottom for?\\\noindent
A: They permit replenishing the tower with fresh air.

\medskip\noindent
Q: If you compare this with the situation on Earth, what does the Sun do in the belt around the equator? What happens with the surface and the air above?\\\noindent
A: The Sun heats the ground which, in turn, heats the air. Just like the model of the updraft tower, the heated air rises and produces a continuous up-current of air.

\medskip\noindent
Q: Can you imagine what happens with the air, when it climbs to high altitudes?\\\noindent
A: The air cools down and moisture condenses to rain.

\medskip\noindent
Q: Coming back to the updraft tower: it had flaps at the bottom to allow replenishing the air. The same happens on Earth. What do we call horizontal air flows?\\\noindent
A: wind

\subsection{Activity 3: Worksheet -- The Wind Engine of the Earth}
We have seen in the experiments that a heat source can heat up the air and cause an upward flow. The very same process happens on Earth.

\medskip\noindent
Q: Look at Fig.~\ref{f:irradiation}. Where on Earth does the Sun heat most efficiently?\\\noindent
A: around the equator

\medskip\noindent
Q: Heating the air directly is quite inefficient. When you think about the updraft tower experiment, the lamp did not heat the air. Describe the process, how eventually the solar energy heats the air.\\\noindent
A: The irradiation from the Sun heats the ground, which in turn heats the air. This is more efficient next to the surface as compared to higher altitudes.

\medskip\noindent
Q: Which part of the atmospheric layers is heated strongest?\\\noindent
A: The one next to the surface.

\medskip\noindent
Indicate the correct attributions:\\\noindent
The air close to the surface of the Earth is warm/cold.\\\noindent
The air at high altitudes above ground is warm/cold.

\medskip\noindent
Q: What happens with the air close to the surface? Consider the temperature differences between low and high altitudes.\\\noindent
A: It rises above the cooler air.

\medskip\noindent
Q: Warm air can store more water than cold air. What happens, when the air rises into higher layers of the atmosphere? Think of boiling water at home, when the hot air meets the cold air or cold surfaces.\\\noindent
A: The water vapour condenses. First to clouds, eventually to larger drops and rain.

\medskip\noindent
Q: Can you explain why the equator regions of the Earth experience so much rain during the year?\\\noindent
A: Warm humid air is driven up to cooler atmospheric layers, where the water condenses to rain. This is a process that works almost all year.

\medskip\noindent
Q: This region around the equator is also called the Intertropical Convergence Zone, abbreviated ITCZ. At some point, the air cannot rise any higher. It is diverted north and south. At those high altitudes, the air constantly cools down. What happens with cold air?\\\noindent
A: Cold air drops closer to the ground.

\medskip\noindent
Q: We just mentioned that cold air can store less water than warm air. Can you explain why the desert areas north and south of the equator are so dry? What happens with the air, when it drops back to the surface?\\\noindent
A: When the warm air rises up to cooler layers, it cannot store water as efficiently and rain falls down. The air is dried by that process.  When this air drops, it is heated up and potentially can store more water. Without replenishment with humid air, the air dries even more.

\medskip\noindent
Q: Back at the surface, the air streams towards the equator region, where they converge (ITCZ). Those are winds we call the Trade Winds. Can you imagine why they are called this?\\\noindent
A: The Trade Winds are a fairly constant phenomenon that helped cargo ships to sail long distances. Its direction is fairly stable as well, so ships didn’t get lost so often.

\medskip\noindent
Q: Try to make a prediction of what would happen with the winds and the weather in general, if the temperature continues to rise.\\\noindent
A: Temperature is the main engine of the updraft. Higher temperatures also lead to more water to be evaporated and enrich the air. As a result, one should expect more extreme phenomena related to the ITCZ.

\medskip\noindent
Q: This is actually happening right now. It is called “Global Warming”. Explain what would happen, if we do not stop this process.\\\noindent
A: This aspect of the climate change will lead to more severe weather, especially near the tropics.

\medskip
We have constructed an entire circulation system that begins and ends at the equator region. This system is called the Hadley cell. Can you draw a schematic with the most relevant elements and processes? Use the prepared sketch below as a starting point.

\begin{figure}[!ht]
\centering
\resizebox{\hsize}{!}{\includegraphics{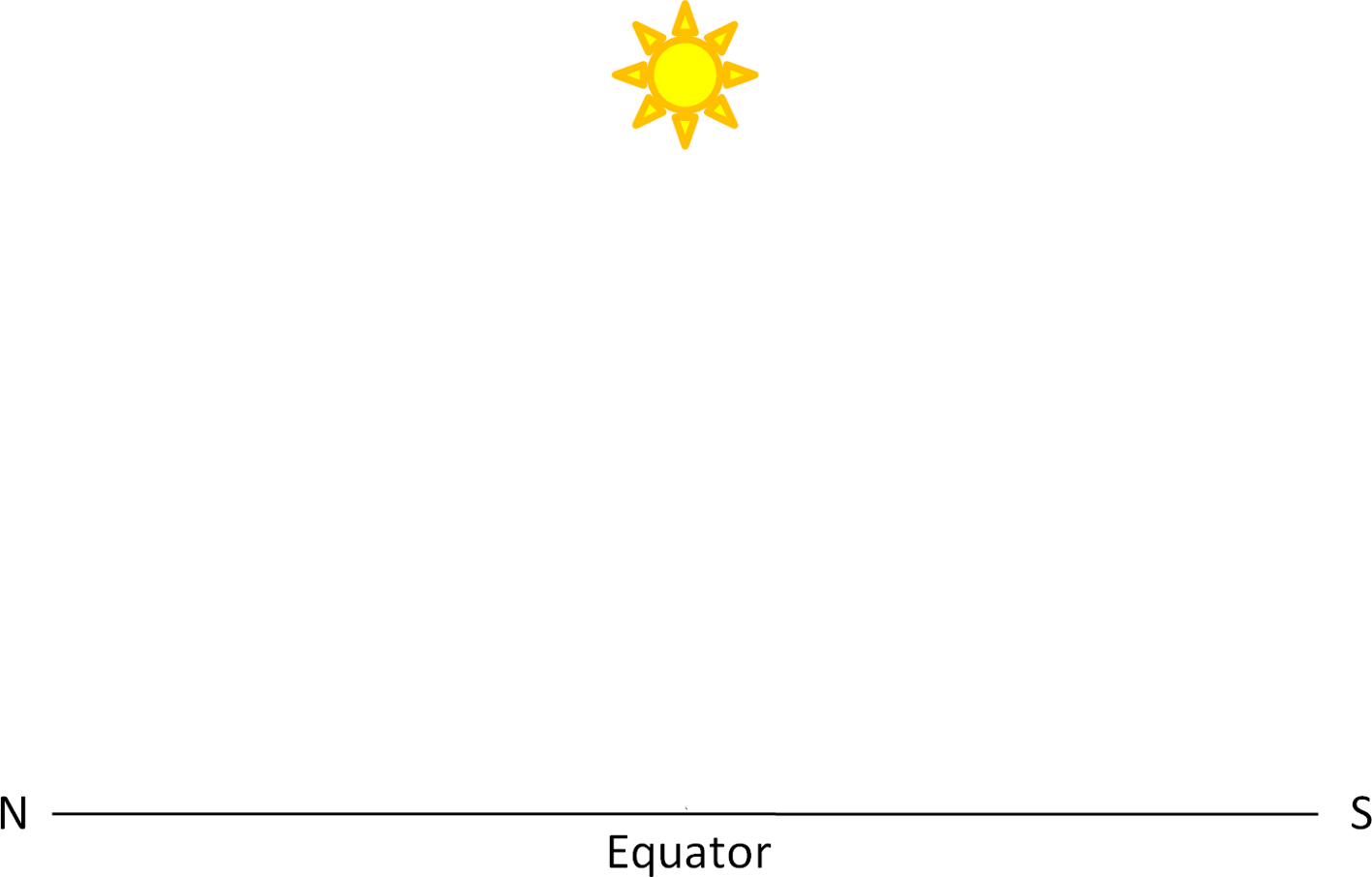}}
\caption{Sketch prepared to draw the Hadley cell (own work).}
\label{f:a3}
\end{figure}

\section{Connection to school curriculum}
This activity is part of the Space Awareness category ``Our Fragile Planet'' and related to the curricula topics:
\begin{itemize}
\item Atmosphere
\item Seasons
\item Composition and structure
\item Surface
\end{itemize}

\subsection{Application to UK curriculum}
The activity covers the following aspects of the subject of Geo\-graphy in the curriculum of the United Kingdom.

\begin{table}[!h]
\begin{tabular}{p{0.05\hsize}p{0.85\hsize}}
\hline\hline
Level & Section \\
\hline
\bf KS2 & \bf Physical geography \\
    & describe and understand key aspects of: physical geo\-graphy, including: climate zones, biomes and vegetation belts, rivers, mountains, volcanoes and earthquakes, and the water cycle\\
\bf KS3 & \bf Locational knowledge \\
    & extend their locational knowledge and deepen their spatial awareness of the world's countries using maps of the world to focus on Africa, Russia, Asia (including China and India), and the Middle East, focusing on their environmental regions, including polar and hot deserts, key physical and human characteristics, countries and major cities \\
\hline
\end{tabular}
\end{table}

\section{Conclusion}
This activity consists of three parts that illustrate the power of the Sun driving a global air circulation system that is also responsible for tropical and subtropical climate zones. Through experiments, students learn how heated air rises above cool air and how a continuous heat source produces air convection streams that can even drive a propeller. Students then apply what they have learnt to complete a worksheet that presents the big picture of the global air circulation system of the equator region to a larger scale.

\begin{acknowledgements}
This resource was developed in the framework of Space Awareness. Space Awareness is funded by the European Commission’s Horizon 2020 Programme under grant agreement no. 638653.
\end{acknowledgements}

%-------------------------------------------------------------------
\bibliographystyle{aa}
\bibliography{Navigation}

\glsaddall
\printglossaries

\begin{appendix}
\section{Relation to other educational materials}
This unit is part of a larger educational package called “Our Fragile Planet” that introduces several historical and modern techniques used for navigation. An overview is provided via: \href{http://www.space-awareness.org/media/activities/attach/6782cf82-139d-4e69-b321-f77312be1fcc/The\%20Climate\%20Box\%20an\%20educational\%20ki\_2udsFBb.pdf}{Our\_Fragile\_Planet.pdf}

\section{Supplemental material}
The supplemental material is available on-line via the Space Awareness project website at \url{http://www.space-awareness.org}. The direct download links are listed as follows:
 
\begin{itemize}
 \item Worksheets: \href{https://drive.google.com/file/d/0Bzo1-KZyHftXZGdhaFZsdWdhRkU/view?usp=sharing}{astroedu1619-Intertropical\_Convergence\_Zone-WS.pdf}
 \item Updraft tower building template: \href{https://drive.google.com/file/d/0Bzo1-KZyHftXeFZrWWxsNWVyQ1k/view?usp=sharing}{astroedu1619-Intertropical\_Convergence\_Zone-ConstructionTemplate.pdf}
 \item Video of working updraft tower: \url{https://youtu.be/rfdQnhONuFs}
\end{itemize}

\end{appendix}
\end{document}